\documentclass[Conference]{IEEEtran}
\IEEEoverridecommandlockouts
\usepackage{cite}
\usepackage{float}
\usepackage{amsmath,amssymb,amsfonts}
\usepackage[ruled,linesnumbered,vlined]{algorithm2e}
\usepackage{algpseudocode}
\usepackage{graphicx}
\usepackage{textcomp}
\usepackage{xcolor}
\usepackage{multirow}
\usepackage{diagbox}
\usepackage{array}
\usepackage{footnote}
\usepackage{makecell} 
\makesavenoteenv{tabular}
\makesavenoteenv{table}
\usepackage{subfigure}
\usepackage{lettrine}
\usepackage{hyperref}
\hypersetup{colorlinks,
citecolor=green,
linkcolor=blue,
urlcolor=black}
\usepackage{pifont}

\begin{document}

\title{User-Centric Machine Learning for Resource Allocation in MPTCP-Enabled Hybrid LiFi and WiFi Networks}

\author{Han Ji\textsuperscript{1}, Declan T. Delaney\textsuperscript{1,}\textsuperscript{2} and Xiping Wu\textsuperscript{1,}\textsuperscript{2}\\

\begin{minipage}{15cm}
\begin{center}
\vspace*{0.3cm}
\textsuperscript{1}{\normalsize\sl School of Electrical and Electronic Engineering, University College Dublin, Dublin, Ireland }\\
\textsuperscript{2}{\normalsize\sl Beijing-Dublin International College, Beijing University of Technology, Beijing, China}\\
{\normalsize  \{han.ji@ucdconnect.ie, declan.delaney@ucd.ie, xiping.wu@ucd.ie\}
\vspace*{-4mm}}\vspace*{0.0cm}\end{center}
\end{minipage}}

\maketitle

\begin{abstract}
As an emerging paradigm of heterogeneous networks (HetNets) towards 6G, the hybrid light fidelity (LiFi) and wireless fidelity (WiFi) networks (HLWNets) have potential to explore the complementary advantages of the optical and radio spectra. Like other cooperation-native HetNets, HLWNets face a crucial load balancing (LB) problem due to the heterogeneity of access points (APs). The existing literature mostly formulates this problem as joint AP selection and resource allocation (RA), presuming that each user equipment (UE) is served by one AP at a time, under the constraint of the traditional transmission control protocol (TCP). In contrast, multipath TCP (MPTCP), which allows for the simultaneous use of multiple APs, can significantly boost the UE's throughput as well as enhancing its network resilience. However, the existing TCP-based LB methods, particularly those aided by machine learning, are not suitable for the MPTCP scenario. In this paper, we discuss the challenges when developing learning-aided LB in MPTCP-enabled HLWNets, and propose a novel user-centric learning model to tackle this tricky problem. Unlike the conventional network-centric learning methods, the proposed method determines the LB solution for a single target UE, rendering low complexity and high flexibility in practical implementations. Results show that the proposed user-centric approach can greatly outperform the network-centric learning method. Against the TCP-based LB method such as game theory, the proposed method can increase the throughput of HLWNets by up to 40\%.

\end{abstract}


\section{Introduction}
\lettrine[loversize=0.1, nindent=0em]{H}{eterogeneous} networks (HetNets) are widely recognised as a key technology towards the sixth generation (6G) communication \cite{wang20236G_road_Review}. As an emerging paradigm of HetNets, hybrid light fidelity (LiFi) and wireless fidelity (LiFi) networks (HLWNets) have attracted an increasing amount of attention in recent years \cite{wu2021hybrid}. In contrast to the HetNets that explore similar spectra, HLWNets are capable of gaining the complementary advantages of the optical and radio spectra. In general, such a hybrid network combines the capacious optical spectra of LiFi and the ubiquitous coverage of WiFi \cite{7402263}. However, due to the heterogeneity of access points (APs), WiFi is prone to suffer from traffic overload than LiFi. Like other forms of HetNets, load balancing (LB) becomes a critical issue that bottlenecks the network capacity of HLWNets.

The existing LB methods for HLWNets \cite{li2015cooperative,wang2017load,ji2022novel,amran2022learning} mostly consider joint AP selection (APS) and resource allocation (RA) problems, presuming that each user equipment (UE) is only connected to one AP, under the constraint of the traditional transmission control protocol (TCP). These methods can be classified into four categories: i) optimisation methods such as \cite{li2015cooperative}; ii) iterative methods such as \cite{wang2017load}; iii) decision-making methods such as \cite{ji2022novel}; and iv) machine learning methods such as \cite{amran2022learning}. Among the above methods, machine learning has been proven to offer a better balance between optimality and computational complexity in comparison with the other types of methods. Recent results show that the implementation time of the learning-aided LB approach for TCP-based HLWNets can be lowered to 100$\mu$s \cite{ji2023adaptive}, which has the potential to meet the ultra-low latency requirement of 6G \cite{you2021towards}.

However, with the traditional TCP, it is infeasible for the UE to enjoy the benefits of different access technologies at the same time. Meanwhile, multipath TCP (MPTCP), which is an ongoing effort of the Internet Engineering Task Force, aims to allow for the simultaneous use of multiple APs. MPTCP is particularly beneficial in the context of HetNets \cite{7864463}, since it can not only gain an aggregated throughput from inverse multiplexing but also add or drop links without disrupting the end-to-end connection, avoiding the annoying handover issue in ultra-dense networks such as LiFi. In \cite{8863828}, an optimisation problem was formulated to tackle the LB task in an MPTCP-enabled HLWNet. However, it only considers two subflows, one from WiFi and one from LiFi. To enable multiple subflows from LiFi, parallel transmission LiFi was proposed in \cite{wu2020parallel}, and the corresponding LB issue was investigated through an iterative approach. Further, upon reinforcement learning, the authors in \cite{10001388} developed a network-centric learning model to adjust the RA coefficients of the subflows in HLWNets. However, this method has two major limitations. First, the method is network-centric, that is it has to make the LB decision for the UEs all together, limiting its practicability. Second, the method aims to maximise the sum throughput with proportional fairness among the subflows, which does not necessarily ensure proportional fairness among the UEs.

The authors' previous work \cite{ji2023adaptive} proposed a user-centric learning model termed as target-condition neural network (TCNN) to handle the APS problem in TCP-based HLWNets. However, there are two main challenges: i) TCP-based TCNN cannot ensure robust connectivity due to possible light path blockage and handover issues since each UE is only served by one AP. This limitation can be tackled by MPTCP. ii) Old TCNN is designed for the APS problem rather than the RA problem, while in MPTCP APS is usually assumed fixed with solutions among a few APs offering the shortest distance. In addition, both the data structure and the learning task vary. In this paper, we tackle these challenges and propose a novel user-centric intelligence-native method to cope with the MPTCP-enabled RA problem in such cooperation-native HLWNets. Results show that the proposed method achieves near-optimal throughput performance within a gap of 7.5\%, while reducing the runtime by over 2,500 times. Compared with the TCP-based LB methods, the proposed approach can increase the network throughput of HLWNets by up to 40\%.

The remainder of this paper is organised as follows. The system model is introduced in Section II. The user-centric learning method is proposed in Section III, including the learning model structure, data collection, training and validation. Simulation results are presented in Section IV. Finally, conclusions are drawn in Section V.

\section{System Model} \label{sec:sys_model}
In this section, the system model of the MPTCP-enabled HLWNet is introduced, including the network topology, the channel model, and the LB problem formulation.

\subsection{Network Topology}      
\begin{figure}[t]
\centering
\includegraphics[width=2.8in]{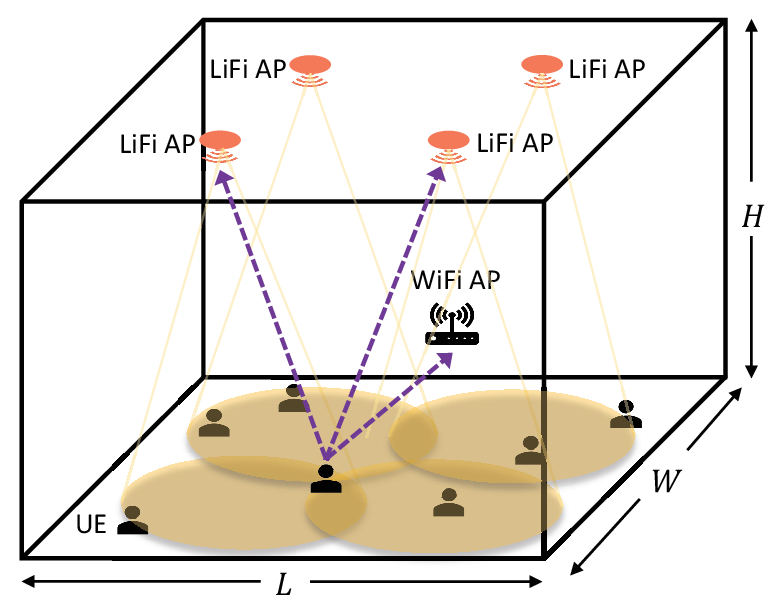} 
\caption{Schematic diagram of an MPTCP-enabled indoor HLWNet.}  
\label{Fig: SystemModel}
\end{figure}

Fig. \ref{Fig: SystemModel} depicts the network topology of the HLWNet that is considered in this paper. Such a hybrid network consists of one WiFi AP and multiple LiFi APs in a room area $L \times W \times H$ (length, width and height). The LiFi APs are arranged in a grid on the ceiling, whereas the WiFi AP is located at the centre of the room at a certain height from the ground. Let $N_a$ and $N_u$ denote the number of APs and the number of UEs, respectively. Let $\mathbb{S}=\{1,2,...,N_a\}$  denote the set of APs, and the set of UEs is $\mathbb{U}=\{1,2,...,N_u\}$. Let $i$ and $j$ denote the index of APs and the index of UEs, where $i\in \mathbb{S}$ and $j\in \mathbb{U}$, respectively. Unlike TCP, MPTCP enables each UE to be served by multiple APs at the same time. Let $N_f$ denote the number of subflows that are employed by one UE. In this paper, each UE adopts $N_f-1$ LiFi links of the best channel quality, in addition to the WiFi link. \mbox{Fig. \ref{Fig: SystemModel}} exemplifies the case of $N_f=3$, where the UE is connected to the WiFi AP and two LiFi APs simultaneously. Wavelength-division multiplexing (WDM) can be adopted to enable the parallel transmission of multiple LiFi links, and the setup details are referred to \cite{wu2020parallel}. Each AP can serve multiple UEs through multiple access techniques, e.g., time-division multiple access, orthogonal frequency-division multiple access, space-division multiple access, etc. Without loss of generality, the aim of this paper is allocating the time resource of each AP. Let $\rho_{i,j}$, which is a continuous variable between 0 and 1, denote the proportion of time resource that AP $i$ gives to UE $j$. 

\subsection{Channel Model} 

The LiFi channel is comprised of the line-of-sight (LoS) and first-order non-line-of-sight (NLoS) paths. The corresponding channel expressions can be found in \cite[eq. (10) and eq. (12)]{kahn1997wireless}. The WiFi channel model can be found in \cite[eq. (7)]{wu2017access}. The signal-to-interference-plus-noise ratio (SINR) of the link between AP $i$ and UE $j$ is denoted by $\gamma_{i,j}$, which is detailed in \cite{wu2020parallel}. The link capacity, which is denoted by $C_{i,j}$, can be expressed as follows:
 \begin{equation}\label{eq:capacity}
C_{i,j} =
\begin{cases}
  \dfrac{{B_i}}{2}{\log_2}\left(1 + {\dfrac{e}{2\pi}} \gamma_{i,j}\right),  & \forall i \in
  {\mathbb{S}}_{\rm{LiFi}}\\ 
  {B_i}{\log_2}(1 + \gamma_{i,j}) , & \forall i \in {\mathbb{S}}_{\rm{WiFi}}  \\ 
\end{cases},
\end{equation}
where $e$ is the Euler’s number; $B_{i}$ denotes the bandwidth of AP $i$; ${\mathbb{S}}_{\rm{LiFi}}$ stands for the set of LiFi APs; ${\mathbb{S}}_{\rm{WiFi}}$ is the set of WiFi APs. The achievable throughput of UE $j$ is denoted by $R_{j}$, which can be computed by:
\begin{equation}\label{eq:UE_throughput}
R_j =  \sum\limits_{i\in{\mathbb{S}_j}} {\rho_{i,j}}C_{i,j},
\end{equation}
where $\mathbb{S}_j$ denotes the set of the APs that are connected with UE $j$. The overall network throughput is $\Gamma =  \sum\limits_{j \in \mathbb{U}}{R_j}$.

\subsection{Load Balancing Problem Formulation}

\begin{figure*}[t]
\centering
\includegraphics[width=6in]{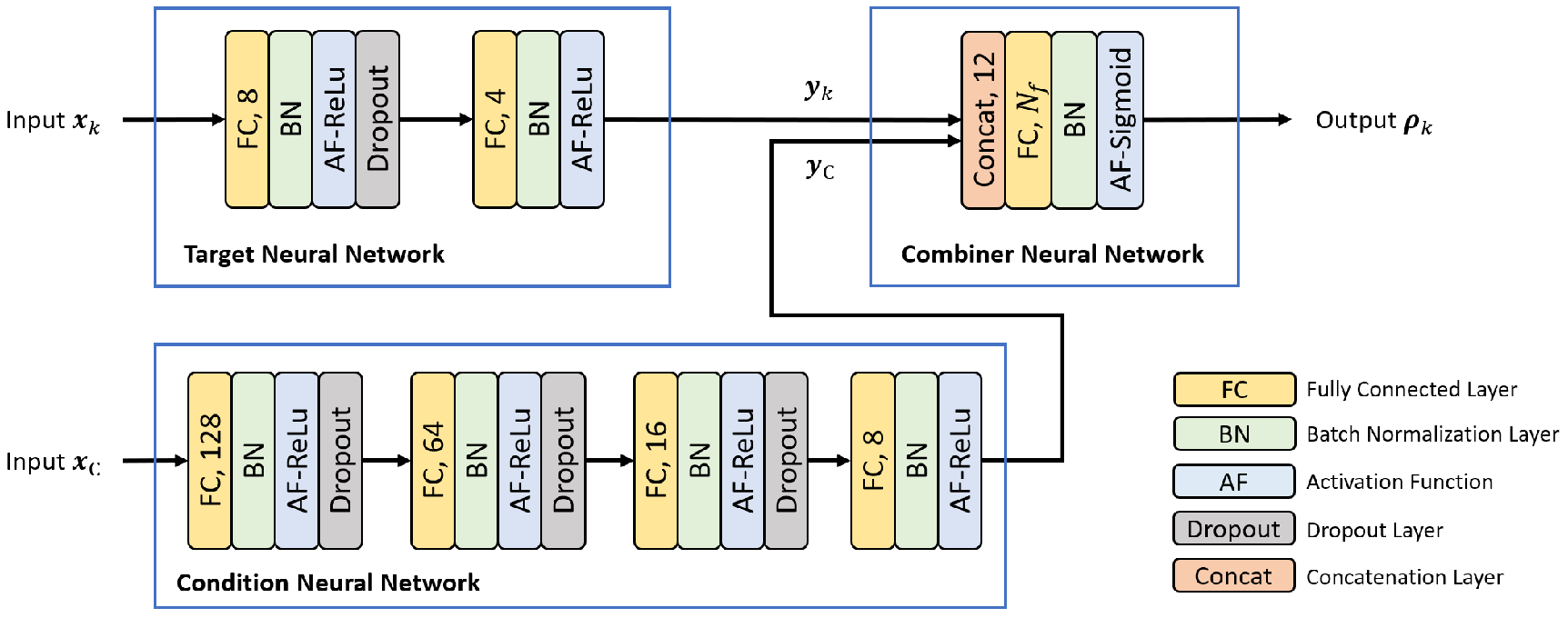} 
\caption{Block diagram of the proposed user-centric learning model.}  
\label{Fig: ATCNN} 
\end{figure*}

The traditional TCP-based LB problem is commonly formulated as a joint optimisation of APS and RA, where the LB functionality is realised through allowing the UE to select an AP that might have a lower channel quality but with a higher resource availability. With MPTCP, this functionality can be achieved by adjusting the traffic loads among the subflows, without disturbing the link connections. Therefore, the LB task in the MPTCP scenario becomes allocating the resources of the APs among their associated subflows, while achieving proportional fairness among the UEs. This problem can be formulated as follows:
\begin{equation}\label{eq:opt}
\begin{array}{l}
\max\limits_{\rho_{i,j}} \;\quad\sum\limits_{j \in \mathbb{U}} {\log}\left({\sum\limits_{i\in{\mathbb{S}_j}}} {\rho_{i,j}}C_{i,j}\right) \\
\;\;{\rm{s}}{\rm{.t}}{\rm{. }}\quad\;\sum\limits_{j\in\mathbb{U}_i}{{\rho _{i,j}} \le 1},  \quad\quad\quad \forall i \in \mathbb{S}; \\
\quad\quad \quad\;\; 0 \le {\rho _{i,j}} \le 1,   \quad\quad\quad \forall\;i,j.\\
\end{array}
\end{equation} 
where $\mathbb{U}_i$ denotes the set of UEs that AP $i$ serves. This nonlinear programming problem can be solved through the OPTI toolbox in MATLAB but requires an excessive amount of computational complexity.



\section{User-Centric Learning Method} \label{sec:proposed_method}

In this section, a novel user-centric deep neural network (DNN) based learning model is developed to tackle the above LB problem, followed by the process of data collection, training and validation. 

\subsection{User-Centric DNN Learning Model}
Fig. \ref{Fig: ATCNN} shows the block diagram of the proposed learning model, inheriting the fundamental structure of TCNN \cite{ji2023adaptive} which involves three key components: one target neural network, one condition neural network, and one combiner neural network. The former two neural networks are used to separately extract the features of the target UE and the other UEs, which are referred to as condition UEs. The outputs of the two neural networks are then fed into the third neural network, which delivers the final output of RA results for the target UE. However, there are two challenges when applying the original TCNN model to cope with the LB issue in the MPTCP scenario. First, the input data in the original TCNN only involves the link quality, whereas the input data here must also contain the link connection status. The additional input would make the learning model more complicate and difficult to train. Second, the original TCNN outputs the link connection status for the target UE, which is a binary classification task since each link is either on or off. In this paper, it becomes a regression task to estimate the resource coefficient for each subflow that is used by the target UE. With the above two challenges addressed, the design of each neural network component is elaborated below.

\subsubsection{Target Neural Network} Let ${\boldsymbol{x}}_k$ denote the input to the target neural network, where $k$ represents the index of the target UE. In the original TCNN, the input is a vector $\boldsymbol{\gamma}_k=\left[\gamma_{1,k}, \gamma_{2,k}, ..., \gamma_{{N_a},k}\right]$ that contains the SNR information on the link between the target UE and each AP, with $N_{a}$ elements in total. As mentioned, in addition to the SNR information, the input here also needs to include the link connection status, indicating which subflows are currently being used for the target UE. Let $\boldsymbol{\chi}_k=\left[\chi_{1,k}, \chi_{2,k}, ..., \chi_{{N_a},k}\right]$ denote the vector of the link connection status, where $\chi_{i,k}=1$ means that the subflow between AP $i$ and the target UE is in use, and otherwise $\chi_{i,k}=0$. The two vectors, $\boldsymbol{\gamma}_k$ and $\boldsymbol{\chi}_k$, can be cascaded into one vector as the input to the target neural network. However, this would double the number of input elements, complicating the learning model. To avoid this issue, the element-wise product of $\boldsymbol{\gamma}_k$ and $\boldsymbol{\chi}_k$ is adopted instead, i.e., ${\boldsymbol{x}}_k = \boldsymbol{\gamma}_k \odot \boldsymbol{\chi}_k$. This operation can retain both the link connection status and the SNR information of the used links while keeping the same input size as in the original TCNN. 

The target neural network is comprised of two fully connected (FC) layers with 8 and 4 neurons, respectively. Each FC layer is followed by a batch normalisation (BN) process to avoid the vanishing gradient problem and a rectified linear unit (ReLU) as the activation function. A dropout procedure with a certain probability, which is denoted by $p$, is adopted before the second FC layer to prevent overfitting \cite{srivastava2014dropout}. The target neural network outputs a latent vector with 4 elements, which is denoted by $\boldsymbol{y}_{k} \in \mathbb{R}^{4}$.

\subsubsection{Condition Neural Network}
The condition neural network is similar to the target neural network, except that the input is about both the target UE and condition UEs. Here the input vector is denoted by $\boldsymbol{x}_{\rm{C}} = [\boldsymbol{x}_1, ..., \boldsymbol{x}_k, ..., \boldsymbol{x}_{N_u}]$. As the input size here is $N_{u}$ times that of the target neural network, four FC layers are adopted to extract the features. The output is another latent vector with 8 elements, which is denoted by $\boldsymbol{y}_{\rm{C}} \in \mathbb{R}^{8}$. 

\subsubsection{Combiner Neural Network}
The above two latent vectors are combined into a third neural network, which employs a concatenation (Cat) layer with 12 neurons, followed by an FC layer and the BN process. The number of neurons in the FC layer depends on the number of subflows used by the target UE, i.e., $N_{f}$. The weight matrix and the bias vector in the FC layer are denoted by $\mathbf{W} \in \mathbb{R}^{12 \times N_f}$ and $\boldsymbol{b} \in \mathbb{R}^{N_f}$, respectively. At last, the sigmoid activation function is used, which is defined as ${f_{\rm{sigmoid}}}(\boldsymbol{x}) = 1 / (1+e^{\boldsymbol{-x}})$. The output of the combiner neural network is $\boldsymbol{\rho}_k \in \mathbb{R}^{N_f}$, which can be computed as follows: 
\begin{equation} 
{\boldsymbol{\rho}_k} = {f_{\rm{sigmoid}}}\left({f_{\rm{BN}}}\left({\mathbf{W}}{f_{\rm{Cat}}}\left(\boldsymbol{y}_{k}, \boldsymbol{y}_{\rm{C}}\right) + \boldsymbol{b}\right)\right),
\end{equation}  
where $f_{\rm{BN}}(\cdot)$ stands for the operation of batch normalisation, and $f_{\rm{Cat}}(\cdot)$ is the operation of concatenation. It is worth noting that the output $\boldsymbol{\rho}_k$ only contains the resource coefficients of the used subflows, of which the original indexes are sorted in an ascending order. For example, if the target UE is using two subflows associated with the 2nd AP and the 7th AP, then the combiner neural network will yield an output ${\boldsymbol{\rho}_k}=[\rho_{1,k},\rho_{2,k}]$, where $\rho_{1,k}$ is the resource coefficient related to the 2nd AP, and $\rho_{2,k}$ is for the 7th AP. This measure can effectively reduce the output size from $N_{a}$ to $N_{f}$, lowering the complexity of the neural network.

\subsection{Dataset Collection}
The sample dataset is collected through numerically solving the LB problem in (\ref{eq:opt}) in a mobile environment. The random waypoint mobility model \cite{RWP_mobility_model} is considered, with a velocity range $[{v_{\rm{min}}}, {v_{\rm{max}}}]$. The simulation period is denoted by $T$, while the sampling period is $T_s$. At each sampling point, the channel state information on both the target UE and the condition UEs (i.e., $\boldsymbol{x}_k$ and $\boldsymbol{x}_{\rm{C}}$) is collected as the input data. The optimal RA solutions provided by (\ref{eq:opt}), which are denoted by $\boldsymbol{\dot{\rho}}_k$, are the ground-truth labels. It is worth noting that the sample dataset differs when the UE number varies. Therefore, the dataset collection as well as the training process is carried out separately for each case of the UE numbers, with $T/T_s$ samples gathered. In addition, the dataset is preprocessed through a linear normalisation and divided into the training dataset and the validation dataset following the 80/20 rule.

\begin{figure}[t] 
\centering
\subfigure[$N_u = 30$.]{\includegraphics[width=0.23\textwidth]{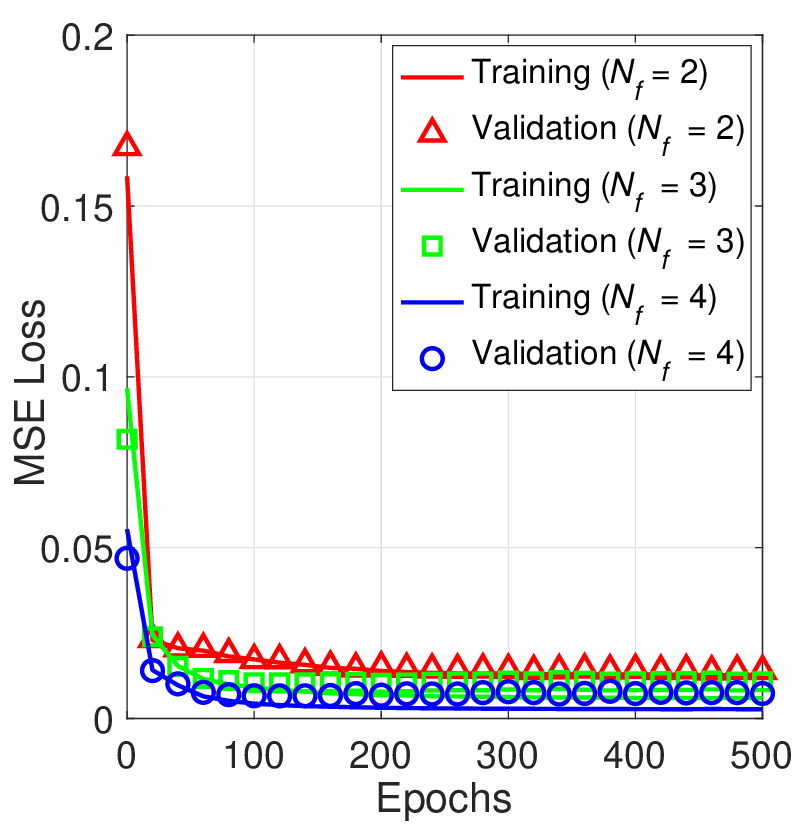} 
\label{fig:Loss_Nf}} 
\subfigure[$N_f = 3$.]{\includegraphics[width=0.23\textwidth]{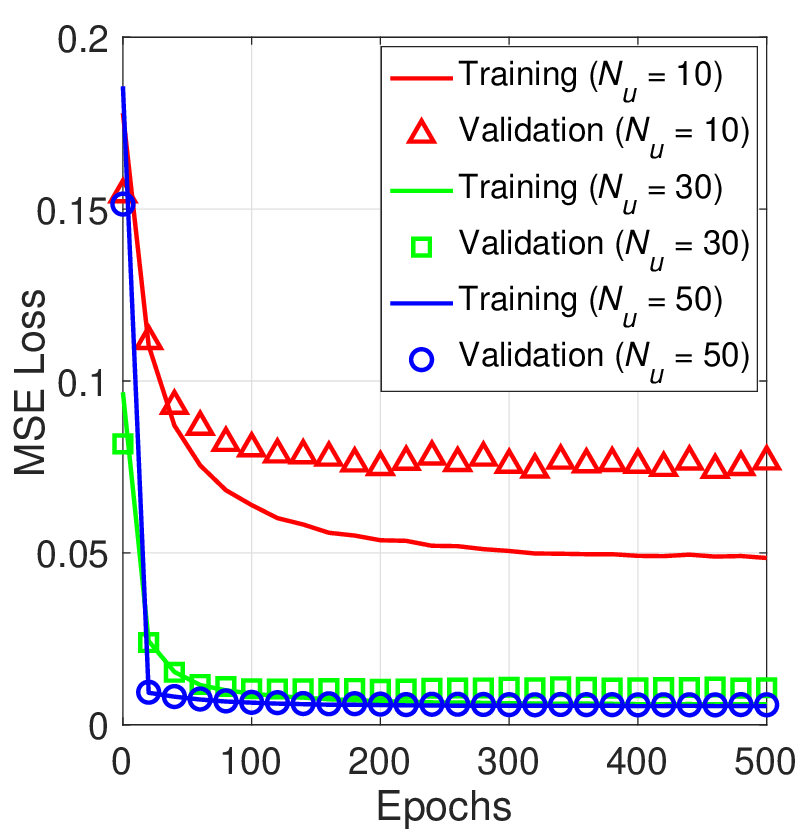}
\label{fig:Loss_Nu}} 
\caption{Training loss and validation loss of the proposed learning model.}
\label{Fig: Training Loss}
\end{figure} 

\subsection{Training and Validation}

Unlike the original TCNN using the cross-entropy loss function to tackle a binary classification task, mean squared error (MSE) is set here to be the loss function, which is commonly used for coping with regression tasks in machine learning. The MSE loss function can be expressed as follows:   
\begin{equation}\label{Equation: MSE Loss} 
L_{\mathrm{MSE}}\left( \alpha \right)= \frac{1}{N}{\sum_{n=1}^{N}}{\left({\boldsymbol{\rho}_{k}^{(n)}} - \boldsymbol{\dot{\rho}}_{k}^{(n)}\right)^2},  
\end{equation}
where $\alpha$ denotes the set of all the weight matrices and bias vectors in the FC layers; $N$ is the number of samples used for training; ${\boldsymbol{\rho}_{k}^{(n)}}$ stands for the estimated resource coefficients for the target UE $k$ in the $n$-th sample; and $\boldsymbol{\dot{\rho}}_{k}^{(n)}$ represents the corresponding ground-truth label. The adaptive moment estimation (Adam) is adopted to train the proposed learning model by iterating $ \alpha \gets \alpha -\eta \nabla L_{\mathrm{MSE}}\left ( \alpha  \right )$, where $\eta$ is the learning rate and $\nabla L_{\mathrm{MSE}}(\alpha)$ stands for the gradient of the loss function with respect to $\alpha$.

Fig. \ref{Fig: Training Loss} presents the training loss and validation loss of the proposed learning model in different cases of $N_{f}$ and $N_{u}$. Fixing $N_{u}$ to be 30 as an example, Fig. \ref{fig:Loss_Nf} shows that the validation loss is very close to the training loss for different subflow numbers, indicating a satisfactory training process with no overfitting or underfitting. Fig. \ref{fig:Loss_Nu} exhibits the training results of different UE numbers, with a fixed $N_{f}=3$. As shown, the validation loss is tightly close to the training loss, except for ${N_u} = 10$, where the validation loss is noticeably higher than the training loss. This is possibly caused by the data sparsity when there are fewer UEs, as a result of the element-wise product operation. However, this does not degrade the throughput optimality of this case in comparison to the other cases. Details will be discussed in Section \ref{sec:sim_throughput}.




\begin{figure*}[h]
\centering   
\subfigure[${N_f} = 2$.]{\includegraphics[width=0.27\textwidth]{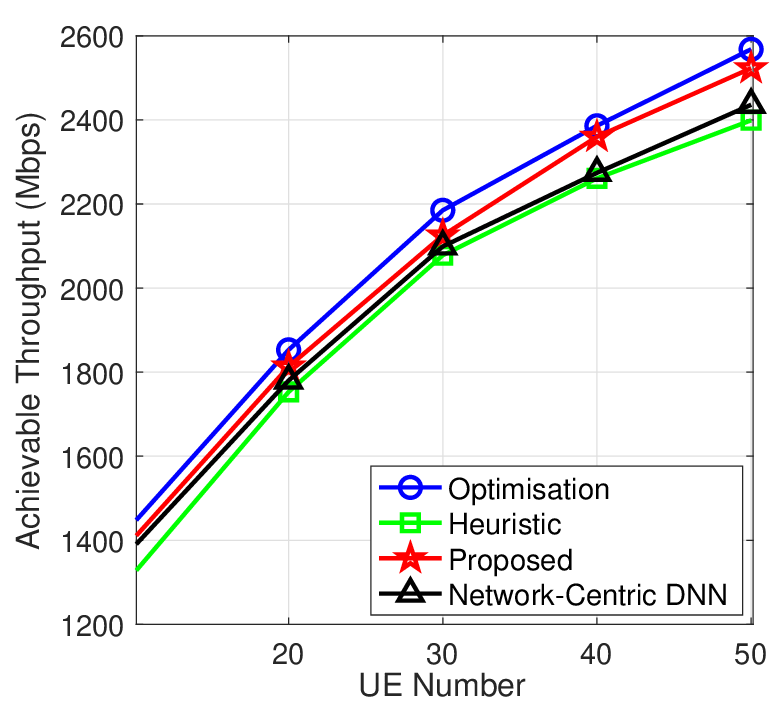} 
\label{fig5:Nf2}} 
\subfigure[${N_f} = 3$.] {\includegraphics[width=0.27\textwidth]{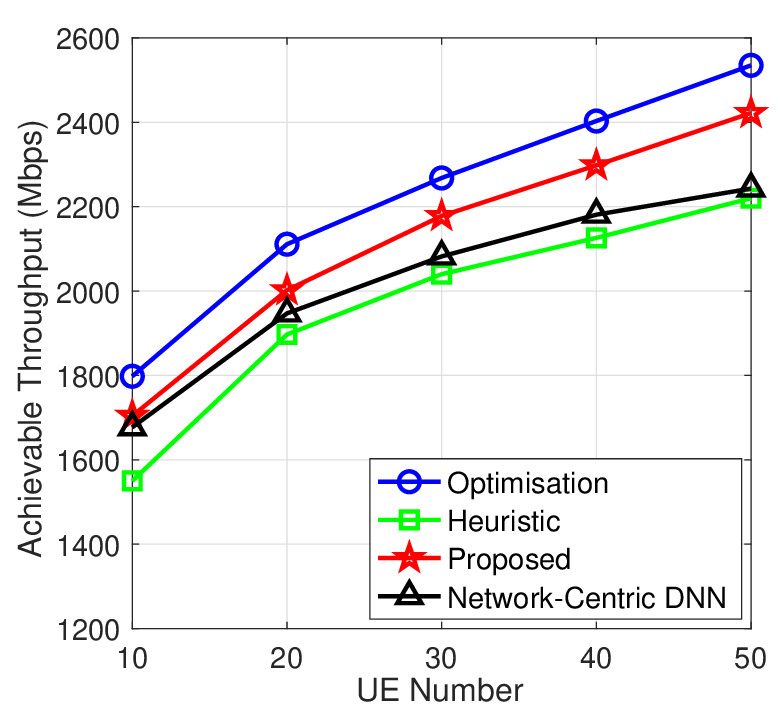}
\label{fig5:Nf3}}
\subfigure[${N_f} = 4$.]{\includegraphics[width=0.27\textwidth]{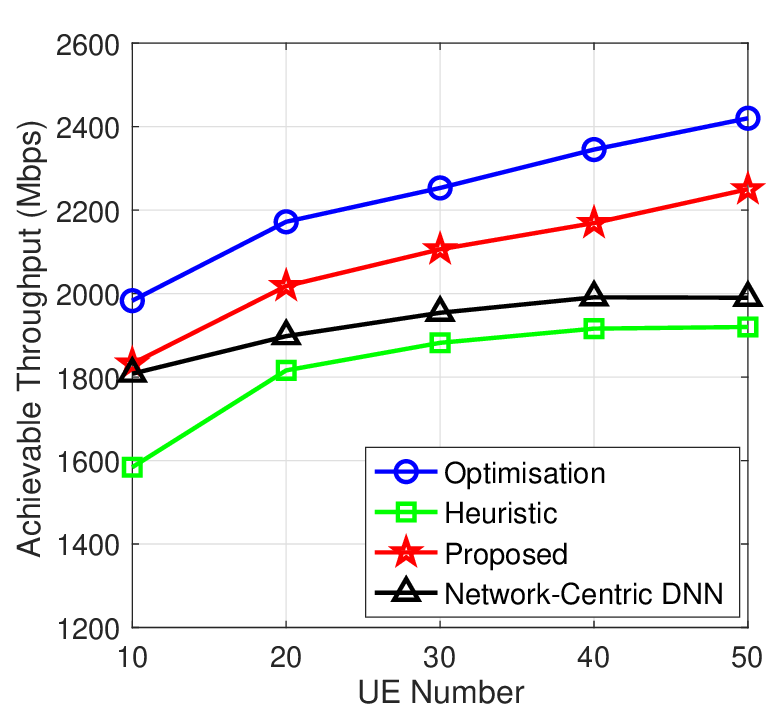} 
\label{fig5:Nf4}} 
\caption{Network throughput of HLWNet with MPTCP.}
\label{Fig: Thr_UEnum}
\end{figure*}


\section{Simulation Results} \label{sec:simulation}

\begin{table}[t]
\renewcommand{\arraystretch}{1.1}
\centering
\caption{Parameter Setup}
\footnotesize
\begin{tabular}{|c!{\vrule width 0.8pt}c|} 
\Xhline{1pt}   
\textbf{Parameters}  & \textbf{Values}  \\  \Xhline{1pt} 
Room size, $L \times W \times H$    & 10m$\times$10m$\times$3m    \\
\Xhline{0.5pt}
Number of LiFi APs  & 16  \\  \Xhline{0.5pt}
Number of WiFi APs  & 1  \\  \Xhline{0.5pt}
LiFi AP separation  & 2.5 m   \\  \Xhline{0.5pt}
UE's height    & 0.5 m   \\  \Xhline{0.5pt}
UE's speed range  & \ $[0.5, 5]$ m/s \\ \Xhline{0.5pt}
Simulation period, $T$  & 500 s   \\ \Xhline{0.5pt}
Sampling period, $T_s$  & 0.1 s   \\ \Xhline{0.5pt}
Dropout probability, $p$ & 0.5   \\ \Xhline{0.5pt}
Learning rate, $\eta$  & 0.001   \\ \Xhline{0.5pt} 
Other LiFi and WiFi parameters   & Refer to \cite{ji2023adaptive} \\
\Xhline{1pt}    
\end{tabular}\label{Table: Parameters}
\end{table}

In this section, Monte Carlo simulations are conducted to evaluate the performance of the proposed user-centric learning model against three baselines in MPTCP: i) the optimisation method in (\ref{eq:opt}); ii) a conventional network-centric DNN model that employs a single neural network to determine the resource coefficients for all the UEs; and iii) a heuristic method that offers proportional fairness to the subflows rather than the UEs. To make a fair comparison, the DNN model contains 5 FC layers. The first four layers use 256, 128, 64 and 32 neurons, respectively, and the last layer adopts $N_u\times N_f$ neurons. Two baselines in TCP are also considered: i) signal strength strategy (SSS), which always connects the UE to the AP that provides the highest channel quality; and ii) game theory (GT) \cite{wang2017load}. Simulations are carried out in Python3.8 on a desk computer with an Intel Core i5-10500@3.1GHz processor. The relevant parameters are summarised in Table \ref{Table: Parameters}. 

\subsection{Network Throughput} \label{sec:sim_throughput}

\subsubsection{MPTCP} Fig. \ref{Fig: Thr_UEnum} presents the network throughput of the proposed method against the noted baselines in MPTCP. In general, compared to the network-centric DNN method, the network throughput achieved by the proposed method is noticeably closer to the optimisation method. Also, the gap between the network-centric DNN and the optimisation method increases as $N_{u}$, while for the proposed method the gap remains almost the same. This signifies that the proposed mechanism of user-centric can provide a higher accuracy than the network-centric learning method. This benefit comes from the attention to different parts of input data in the user-centric structure. Further, it is found that employing more subflows can increase the network throughput when $N_{u}$ is less than 30, but it reduces the throughput for a larger $N_{u}$. The reason behind this trend is two-fold. On the one hand, the resource competition is less fierce when $N_{u}$ is smaller, and using more subflows can aggregate the available resource of different APs for the same UE, boosting its throughput. On the other hand, the resource competition becomes more intense when $N_{u}$ is larger. In this case, using more subflows allows a more flexible way to achieve proportional fairness among the UEs, at the cost of sacrificing the overall throughput to some extent. See a detailed analysis of the UE fairness in Section \ref{sec:fairness}.


\begin{figure}[t]
\centering
\includegraphics[height=2.4in,width=2.8in]
{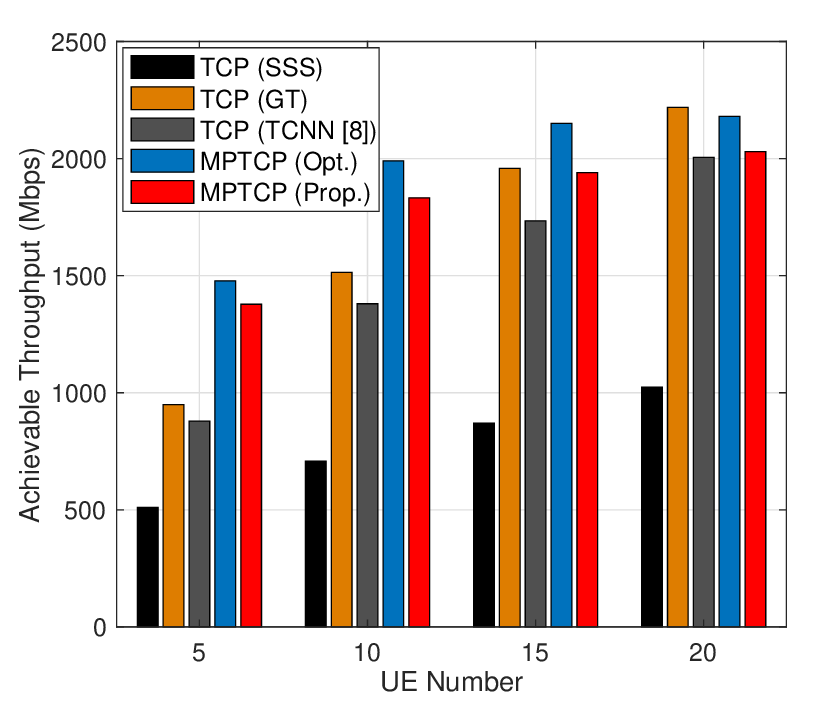} 
\caption{Network throughput comparison between TCP (including SSS, GT and the traditional TCNN \cite{ji2023adaptive}) and MPTCP (including optimisation and the proposed learning method), where $N_f = 4$.}  
\label{fig5:TCP}
\end{figure}

\subsubsection{MPTCP versus TCP} Fig. \ref{fig5:TCP} evaluates the throughput performance of the proposed method against the baselines in TCP. Here four values of $N_{u}$ are involved: 5, 10, 15, and 20 and $N_{f}=4$ is presented for MPTCP. As shown, two key points can be concluded: i) an inevitable gap (less than 10\%) exists between the learning method and the optimal solution in both TCP and MPTCP scenarios; ii) MPTCP brings higher throughput than TCP for both optimisation and learning methods. For the first point, this is because the training dataset used in learning models is collected via the optimisation method, causing a marginal performance gap. For the second point, the benefit can be attributed to the usage of multiple paths, especially for smaller UE numbers. For example when $N_u = 5$, the proposed MPTCP-based learning model can increase the 41\% and 57\% throughput of the GT and TCP-based TCNN methods, respectively. When $N_{u}$ reaches 20, the TCP-based TCNN method obtains close throughput as the proposed method in MPTCP. This is because the usage of resources tends to be saturated as $N_{u}$ increases, weakening the benefit of MPTCP.
However, it is worth noting that GT and optimisation methods require a substantial amount of runtime, ranging from several milliseconds to several hundred milliseconds \cite{ji2023adaptive}. In contrast, the proposed method just needs a runtime less than 100$\mu$s. See details in Section \ref{sec:runtime}.


\subsection{UE Fairness} \label{sec:fairness}
\begin{figure}[t]
\centering
\includegraphics[height=2.5in,width=2.8in] 
{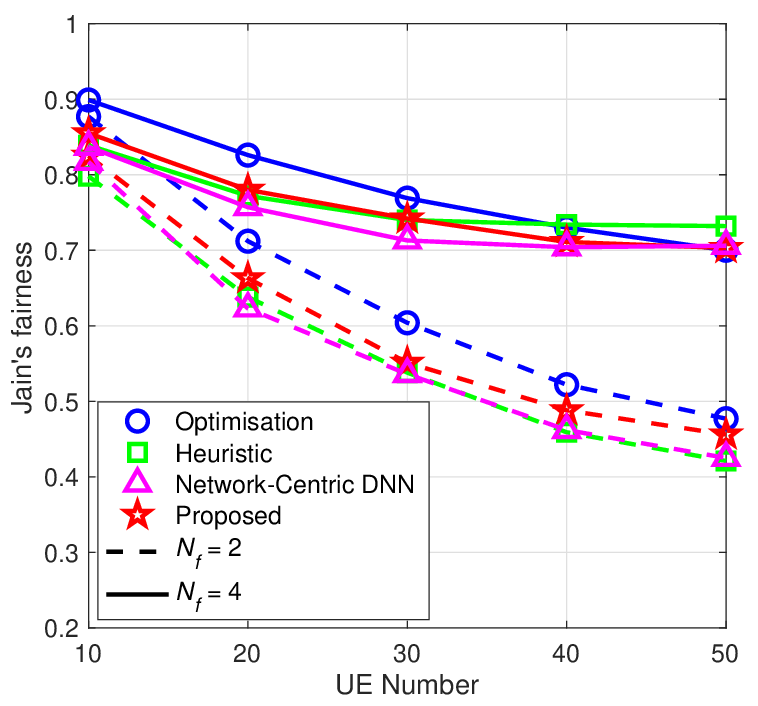} 
\caption{UE fairness versus the number of UEs.}  
\label{Fig: Fairness}
\end{figure}

The Jain's fairness index \cite[eq. (3)]{ji2023adaptive} is measured in Fig. \ref{Fig: Fairness}. In general, the UE fairness reduces as $N_{u}$ increases, due to the more fierce resource competition. It can also be observed that the UE fairness increases as $N_f$, especially for a larger $N_{u}$, as explained. Further, the proposed user-centric learning method offers a higher UE fairness than the network-centric DNN. Compared with the optimisation method, the proposed method exhibits a fairness reduction between 0.02 and 0.06 when $N_{f}=2$. This gap diminishes as $N_{u}$ or $N_{f}$ increases. When $N_u=50$ and $N_f=4$, the proposed method achieves almost the same UE fairness as the optimisation method.

\subsection{Inference Time} \label{sec:runtime}

The inference time of the proposed learning model is summarised in Table \ref{Table: runtime}, in comparison with the baselines in MPTCP. The optimisation method takes the longest inference time, as expected. In addition, the inference time of the optimisation method linearly increases with $N_{u}$, in line with the fact that the corresponding Big-O complexity is $\mathcal{O}(N_{a}N_{u})$. Similarly, a noticable increase in the runtime is observed in the case of network-centric DNN, because its input and output sizes are both proportional to $N_{u}$. In contrast, the inference time of the proposed method exhibits a very slight increase with $N_{u}$. This is because within the proposed user-centric learning model, only the size of the condition neural network is proportional to $N_{u}$, while the sizes of the other two neural networks are independent of $N_{u}$. In summary, the proposed method is implemented faster than the network-centric DNN. When $N_u = 50$ and $N_f = 2$, for example, the optimisation method and the network-centric DNN method cost 350ms and 96$\mu$s, respectively, whereas the proposed model just needs 65$\mu$s, which is 5,300 times as fast as the optimisation method. 

\section{Conclusion} \label{sec:conclusion}
In this paper, a user-centric learning method was proposed to tackle the resource managment problem for HLWNets in an MPTCP scenario. Unlike the conventional network-centric method which determines the resource coefficients for the subflows belonging to all the UEs, the proposed method outputs the resource coefficients for the subflows of a single target UE. This is realised by the unique structure of target-condition neural network, which handles resource allocation for the target UE upon the information on the other UEs. Results show that compared to the network-centric DNN method, the proposed user-centric method can achieve a network throughput closer to the optimal solution, with reduced inference time. Against the TCP-based LB method such as game theory, the proposed method can increase the network throughput of HLWNets by up to 40\%, which is attributed to the benefit of MPTCP.

\begin{table}[t]
\centering
\renewcommand{\arraystretch}{1.1}
\setlength\tabcolsep{1.2 mm}
\footnotesize
\caption{Comparison of Inference Time (in Milliseconds).}
\begin{tabular}{|c|c|c|c|c|c|c|}
\hline 
\multicolumn{2}{|c|}{\diagbox[]{Method} {UE number}} & 10 & 20 & 30 & 40 & 50    \\ 
\hline
\multirow{4}{*}{\vtop{\hbox{$N_f = 2$}}} & Optimisation & 137.9 & 176.5 & 203.7 & 301.5 & 350.8   \\ 
\cline{2-7}
    & Heuristic & 0.090 & 0.095 & 0.096 & 0.103 & 0.097    \\  
\cline{2-7}
    & Network-centric DNN & 0.065 & 0.073 & 0.084 & 0.087 & 0.096 \\       
\cline{2-7}
    & \textbf{Proposed} & \textbf{0.049} & \textbf{0.052} & \textbf{0.060} & \textbf{0.063} & \textbf{0.065}  \\
\cline{2-7}
\hline
\multirow{4}{*}{\vtop{\hbox{$N_f = 3$}}} & Optimisation & 132.7 & 182.7 & 207.4 & 244.8 & 276.9  \\ 
\cline{2-7}
    & Heuristic & 0.086 & 0.098 & 0.113 & 0.113 & 0.114   \\ 
\cline{2-7}
    & Network-centric DNN & 0.071 & 0.079 & 0.087 & 0.095 & 0.097 \\
\cline{2-7}
    & \textbf{Proposed} & \textbf{0.048} & \textbf{0.053} & \textbf{0.057} & \textbf{0.059} & \textbf{0.067}  \\
\cline{2-7}
\hline
\multirow{4}{*}{\vtop{\hbox{$N_f = 4$}}} & Optimisation & 117.3 & 160.8 & 191.3 & 243.0 & 273.4 \\ 
\cline{2-7}
    & Heuristic & 0.103 & 0.102 & 0.100 & 0.112 & 0.105 \\  
\cline{2-7}
    & Network-centric DNN & 0.073 & 0.081 & 0.087 & 0.096 & 0.103 \\
\cline{2-7}
    & \textbf{Proposed} & \textbf{0.051} & \textbf{0.054} & \textbf{0.061} & \textbf{0.062} & \textbf{0.066} \\
\cline{2-7}
\hline
\end{tabular}\label{Table: runtime}
\end{table}

\section*{Acknowledgment}
This work is funded by Beijing-Dublin International
College, a joint international college between University College Dublin and Beijing University of Technology. Han Ji acknowledges support from the China Scholarship Council Grant.

\bibliographystyle{IEEEtran}
\bibliography{IEEEabrv, Reference}

\end{document}